\documentclass[aps,prb,portrait,twocolumn,superscriptaddress]{revtex4-1}

\usepackage{graphicx}
\usepackage{dcolumn}
\usepackage{bm}
\usepackage{longtable}
\usepackage{epsfig}
\usepackage{amsmath}

\begin{document}

\title{A Bespoke Single-Band Hubbard Model Material}

\author{S. M. Griffin}
\affiliation{Materials Theory, ETH Zurich, Wolfgang-Pauli-Strasse 29, CH-8093 Z\"urich, Switzerland}
\author{P. Staar}
\affiliation{Institute for Theoretical Physics, ETH Zurich, CH-8093 Z\"urich, Switzerland}
\author{T. C. Schulthess}
\affiliation{Institute for Theoretical Physics, ETH Zurich, CH-8093 Z\"urich, Switzerland}
\affiliation{Swiss National Supercomputing Center, ETH Zurich, CH-6900 Lugano, Switzerland}
\author{M. Troyer}
\affiliation{Institute for Theoretical Physics, ETH Zurich, CH-8093 Z\"urich, Switzerland}
\author{N. A. Spaldin}
\email{nicola.spaldin@mat.ethz.ch}
\affiliation{Materials Theory, ETH Zurich, Wolfgang-Pauli-Strasse 29, CH-8093 Z\"urich, Switzerland}

\begin{abstract}
The Hubbard model, which augments independent-electron band theory with a single parameter to describe
electron-electron correlations, is widely regarded to be the `standard model' of condensed matter physics. 
The model has been remarkably successful at addressing a range of correlation effects in solids, but beyond one 
dimension its solution is intractable. Much current research aims, therefore, at finding appropriate approximations 
to the Hubbard model phase diagram. Here we take the new approach of using \textit{ab initio} 
electronic structure methods to design a material whose Hamiltonian is that of the single-band Hubbard model.
Solution of the Hubbard model will then be available through measurement of the material's properties. 
After identifying an appropriate crystal class and several appropriate chemistries, we use density functional 
theory and dynamical mean-field theory to screen for the desired electronic band structure and metal-insulator 
transition. We then explore the most promising candidates for structural stability and suitability for doping
and propose specific materials for subsequent synthesis. Finally, we identify a regime -- that should manifest 
in our bespoke material -- in which the single-band Hubbard model on a triangular lattice exhibits exotic $d$-wave 
superconductivity.
\end{abstract}

\maketitle

\section{Introduction}

Independent-electron band theory -- in which interactions between electrons are treated at a mean-field
level -- is a simple yet often remarkably effective method for calculating the behavior of electrons in solids. 
In its modern density functional implementation, its combination of explicit description of crystal structure
and chemistry with computational affordability has led to many successes ranging from early predictions of
phase stability and lattice dynamics in technologically important materials such as semiconducting silicon\cite{Yin/Cohen:1982}, 
through description of the structure and electronics of lattice defects\cite{Freysoldt_et_al:2014}, to modern predictions
of new physics and phenomena in complex materials such as nano- and heterostructures\cite{Rondinelli/Spaldin:2011}. 

\begin{figure*}
 \centering
 \includegraphics[width=4cm, keepaspectratio=true]{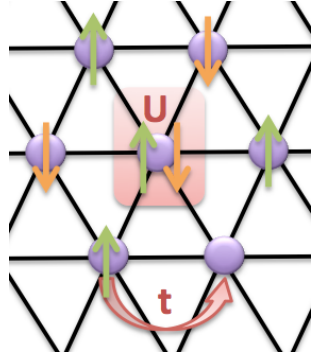}
 \caption{Cartoon of a triangular lattice illustrating the two parameters that enter the Hubbard model: the inter-site hopping, $t$, and 
the onsite repulsion, $U$.} 
 \label{hopping}
\end{figure*}

By construction, however, independent-electron theories do not include explicit correlations between individual
electrons and so are often inappropriate for describing or predicting the properties of so-called strongly-correlated
materials, in which these interactions dominate the physics. Its classic failure is the inability to predict an
insulating state for a half-filled band, which will always be metallic within the conventional band theory formalism. 
Here model Hamiltonians with explicit electon-electron
interaction terms have an advantage, at the expense of loss of computational tractability and chemical information.
In particular, Hubbard introduced the modern form of the model that now bears his name in a series of papers beginning in 
1963\cite{Hubbard:1963, Hubbard:1964, Hubbard2:1964} specifically to ``set up the simplest possible model containing 
the necessary ingredients''.  
(We note that the first hints of the Hubbard model can be found in the quantum chemistry literature in the 
1950s\cite{PariserI:1953, PariserII:1953, Pople:1953}, with P. W. Anderson also suggesting an early variation\cite{Anderson:1959},
and the standard form also independently proposed by Gutzwiller\cite{Gutzwiller:1963} and Kanamori\cite{Kanamori:1963}). 
The Hubbard Hamiltonian is
\begin{equation}
 H=-t\sum_{\langle i,j\rangle \sigma}(c^{\dagger}_{i,\sigma}c_{j,\sigma}+
c^{\dagger}_{j,\sigma}c_{i,\sigma})+U\sum^{N}_{i=1}n_{i\uparrow}n_{i\downarrow}
\quad .
\label{hubbard}
\end{equation}
Here the first term is the kinetic energy term of standard band theory, with $\langle i,j\rangle$ indicating summation 
over all pairs of lattice sites (usually nearest neighbours), $\sigma$ indicating the spin degree of freedom, and $t$ the
hopping matrix element between states $i$ and $j$. $c^{\dagger}_{i}$ and $c_{i}$ are the usual creation and annihilation 
operators, and the number operator, $n_i = c^{\dagger}_{i}c_{i}$.
The second term introduces explicit electron-electron correlations through the Coulomb repulsion, of magnitude $U$, 
between two electrons occupying the same lattice site. We illustrate both terms schematically in Fig.~\ref{hopping}. 

For small values of $U$ and large values of $t$ (the weak-correlation limit), Eqn.~\ref{hubbard} yields metallic solutions 
for partially filled bands and can successfully describe, for example, itinerant magnetism in transition metals. 
Importantly, increasing the ratio of $U/t$ (the strong-correlation limit) causes a crossover 
to an insulating solution, allowing the description of, for example, the antiferromagnetic insulating state of transition 
metal oxides. Modern studies can incorporate (non-self-consistent) information about crystal structure and chemistry via 
density functional calculations of the $t$ and $U$ parameters, and an increasing range of materials from Mott insulators 
to high-$T_{c}$ superconductors is currently addressed\cite{Nekrasov_et_al:2000, Held_et_al:2001,Lichtenstein/Katsnelson:2000, Weber_et_al:2010}.
While it is not obvious why a single-band picture is relevant in the case of systems with degenerate $d$ orbitals or strong hybridization with $s$ or $p$ states, 
several works show that the single-band picture does indeed successfully describe low-energy charge and spin excitations in a realistic manner\cite{Zhang/Rice:1988, Luo/Bickers:1993}.

Despite the simplicity of Eqn.~\ref{hubbard}, the general Hubbard Model is in fact computationally intractable, and 
much current research activity aims at obtaining the Hubbard model phase diagram. Exact solutions are possible 
in one dimension\cite{Essler:2005}, where a transition from a metal to a Mott insulator occurs at half filling when 
$U>t$\cite{Lieb/Wu:1968}. Away from exactly half filling the solution is always metallic for either hole or electron 
doping. 
In two dimensions, numerically exact solutions can be found in the ground state only for either very small lattices of 
at most about twenty sites \cite{1559996}, or on so-called ladder models consisting of coupled chains for around 100 sites
using the density matrix renormalization group algorithm\cite{noack1997}.  For arbitrary dimension and large lattices, however, approximations must be
used to investigate the phase diagram.
Analytical methods such as slave-boson mean field theory and the Gutzwiller approximation have been applied 
extensively\cite{Kotliar/Ruckenstein:1986, Hasegawa:1997, Bunemann/Gebhard:2007}, and  
numerical techniques such as the Dynamical Cluster Approximation (DCA) and Dynamical Mean-Field Theory (DMFT) have 
had some success\cite{Maier_et_al:2005, Koga_et_al:2004, Held_et_al:2001,Gull:2013hhb}. The rapid scaling of the computational resources required with the system size has hampered 
efforts in simulating large systems, however. 
Despite this profusion of computational approaches, consensus has yet to be reached on 
the exact nature of the phase diagram throughout the entire phase space, although diverse numerical methods start to agree on some properties of the 2D Hubbard model \cite{Scalapino:2007,1505.02290}. Much less is known in three dimensions, where substantial interest has been raised by the possibility of realizing the Hubbard model in optical lattices \cite{Jaksch200552,Esslinger:2010ex,Lewenstein:2007p32635} and simulations have mainly been performed by quantum Monte Carlo simulations at relatively high temperatures of the order of the N\'eel temperature \cite{Staudt:vz,Jarrell:2005ec,Fuchs:2011ch,PhysRevLett.107.086401,Kozik:2013nt,Imriska:2014dg}.

In this work we take a new approach to ``solving'' the single-band Hubbard model and use \textit{ab initio} electronic 
structure methods, at the band theory and dynamical mean field theory levels, to design a real material whose Hamiltonian 
is exactly that of the single-band Hubbard model. Measurement of the properties of our bespoke material would then allow 
extraction of the Hubbard model phase diagram. 
Our approach is somewhat reminiscent of experimental simulations of the
Hubbard model that have been performed using ultracold atoms\cite{Greiner:2002}, which take the role of the 
the electrons in the conventional Hubbard model. The atoms are trapped in a potential created by interfering laser 
beams and since the 
lasers are highly controllable, a vast range of potentials can be created using different interference effects. 
The simulated lattice model is therefore highly tunable both in terms of the lattice geometry and dimensionality, and in the 
value of $t/U$. 
In particular, a metal-insulator transition, accompanied by a suppression of doubly-occupied sites,
a drop in compressibility and an excitation gap, has been demonstrated in a gas of cold fermionic atoms 
by tuning the strength of their mutual repulsion \cite{Jordens:2008, Schneider:2008is}, and  
short range magnetic correlations have been observed \cite{Greif:2013kb,Hart:2015ex}. 
While proposals have been made to cool below the N\'eel temperature, however, (see, for example Ref.~\onlinecite{PhysRevLett.107.086401}) 
to achieve long range magnetic oder and adiabatically prepare $d$-wave superconducting phases \cite{Trebst:2006ga}, experimental progress 
has been slow and challenging. A condensed matter realization of the Hubbard model, even if not as tunable as ultracold quantum gases, 
will thus be interesting as it will allow much lower temperatures to be reached.

\section{Design of a single-band Hubbard model material}

\subsection{Candidate crystal structures and chemistries}

\begin{figure*}
 \centering
 \includegraphics[width=16cm, keepaspectratio=true]{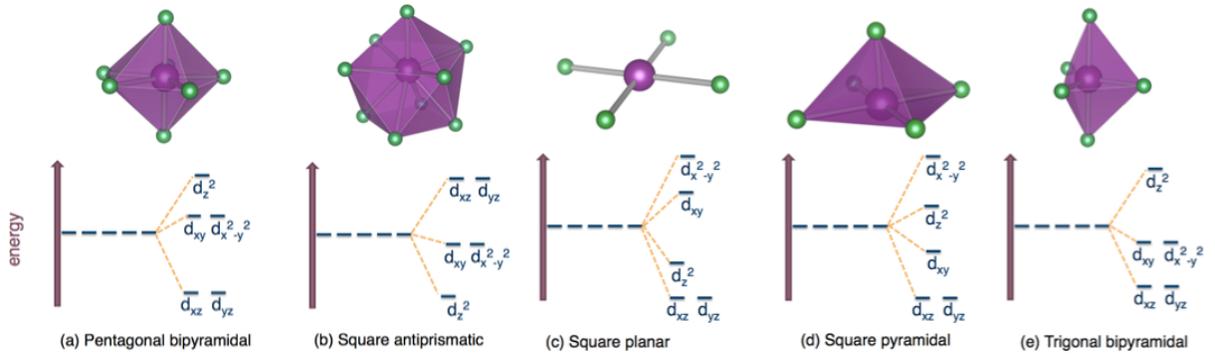}
 \caption{Common coordination environments for transition-metal ions that yield at least one singly degenerate crystal-field
level. (a) pentagonal bipyramidal, (b) square antiprismatic, (c) square planar, (d) square pyramidal and (e) trigonal bipyramidal.}
 \label{crystalfields5}
\end{figure*}

Since we seek a material described by a {\it single-band} Hubbard model, we begin by engineering a non-degenerate 
$d$-manifold using crystal-field 
considerations. Crystal-field splittings with single non-degenerate $d$ bands are obtained for a variety of ionic 
geometries, such
as pentagonal bipyramidal, square antiprismatic, square planar, square pyramidal and trigonal bipyramidal 
(Fig.~\ref{crystalfields5}). In this work we 
choose the trigonal bipyramidal coordination, in which the five-fold degeneracy is broken into two doubly-degenerate 
orbital levels ($d_{xz}, d_{yx}$ and $d_{xy}, d_{x^2-y^2}$), and a single orbital ($d_{z^2}$), for its combination 
of large energetic separation of the single-band state and geometric simplicity. For trigonal bipyramidal coordination 
with shorter apical than in-plane ligand distances, the isolated $d_z^2$ band is the highest energy level, and so it 
becomes half filled for a $d^9$ electron configuration.

\begin{figure*}
 \centering
 \includegraphics[width=12cm, keepaspectratio=true]{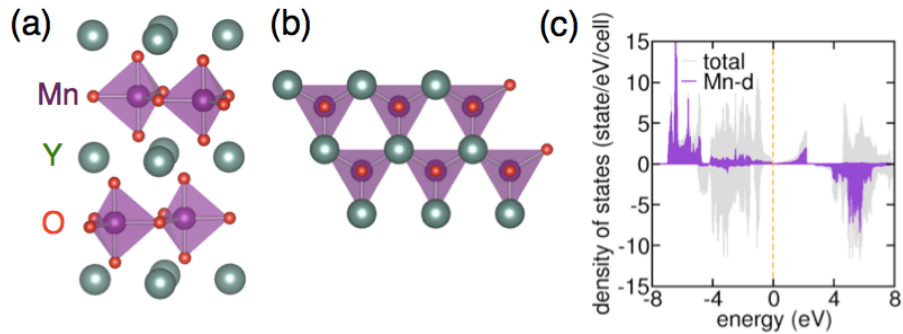}
 \caption{(a) Side view of the hexagonal-manganite structure of YMnO$_3$. Layers of corner-shared MnO$_5$ trigonal bipyramids are 
separated  by layers of Y ions. 
(b) Top view of the hexagonal-manganite structure showing the triangular lattice formed by the Mn ions which are 
connected via shared oxygen ions. 
(c) Calculated density of states of hexagonal manganite YMnO$_{3}$ in its high symmetry reference structure with the
Mn ions ordered ferromagnetically. The orbitally-projected Mn-$d_{z}^{2}$ states are shaded in purple and the Fermi
energy, E$_{F}$, is set to $0$ eV. Majority (minority) spin states are shown on the positive (negative) $y$ axis. }
 \label{ymo_structure}
\end{figure*}

Such trigonal bipyramidal coordination is found in the well-studied hexagonal manganite structure adopted by YMnO$_3$ as
well as the manganites of the smaller rare earth ions. The structure is formed from layers of corner-sharing MnO$_5$ trigonal 
bipyramids alternating with planes of Y ions. Figs.~\ref{ymo_structure}(a) and (b) show the high-symmetry $P6_{3}/mmc$ variant that 
occurs at high temperature; below around 1000K, there is a phase transition to a ferroelectric $P6_{3}cm$ structure, which does not 
alter the coordination environment and hence does not change the crystal-field splitting. 
In Fig.~\ref{ymo_structure}(c) we show our calculated density of states for hexagonal YMnO$_{3}$ in the high symmetry
$P6_{3}/mmc$  reference structure, and with the local moments on the Mn ions ferromagnetically aligned. The $3d^4$ Mn$^{3+}$ 
ions have a high-spin configuration, with the up-spin $d_{xz}, d_{yx}$ and $d_{xy}$ and $d_{x^2-y^2}$ orbitals occupied, and the spin-up 
$d_{z^2}$ forming the lowest energy state above the Fermi level. The exchange-split spin-down states begin at an energy of  
$\sim$4 eV above the Fermi level. While the $d^{4}$ Mn$^{3+}$ configuration is magnetic and does not yield a half-filled single band, 
excluding it from consideration as our single-band Hubbard material, we see already that the $d_z^2$ orbitals form an isolated band 
with little hybridization with the oxygen ligands, confirming our intuition that this crystal class is promising. 

Next we select ion combinations that have formal valence states that should yield half-filling of the isolated $d_z^2$ band, and will
also likely allow for doping across the range from empty to filled band. As mentioned above, to half-fill the $d_{z^2}$ band, our B-site
cation should have a $d^9$ configuration, suggesting Cu$^{2+}$ as the most promising candidate. For oxides or sulphides, both of which
have divalent anions, the A-site cations should be four valent suggesting Zr or Sn as possibilities. For fluorides, in which the anions
are monovalent, the A-sites should also be monovalent, suggesting Li or Na. We take, therefore the following as our initial list of
trial compounds for further study: ZrCuO$_3$, SnCuO$_3$; ZrCuS$_3$, SnCuS$_3$; LiCuF$_3$, NaCuF$_3$.

\subsection{Computational details}
Our electronic structure calculations were performed within density-Functional theory using the Vienna \textit{ab initio}
Simulation Package (VASP) \cite{VASP1, VASP2}. The electronic wavefunctions and density were expanded using a plane-wave basis
set, and we used projector-augmented-wave (PAW) potentials \cite{PAW} for core-valence separation. For the exchange-correlation 
potential we used the generalized gradient approximation (GGA)\cite{PBE1, PBE2} and strong correlation effects were treated by 
means of the GGA+$U$ scheme. Here we used the Dudarev method\cite{Dudarev_et_al:1998} in which a $U_{eff}=U-J$ accounts for the 
on-site $U$ (Coulomb replusion) and $J$ (Hund's rule exchange) on the metal $d$ states. To choose an appropriate $U_{eff}$ for 
hypothetical compounds (without spectroscopic data), we performed hybrid functional calculations to select an appropriate range 
by matching the DFT+$U$ band gap with the hybrid functional result. In the hybrid functional calculations, an HSE functional 
consisting of 75\% PBE and 25\% exact Hartree-Fock exchange was used\cite{hse}. For our Cu-based compounds, this led us to choose 
a $U_{eff}$ of 7 eV on the  Cu-$d$ states, which is similar to that used in previous studies on cuprate 
materials\cite{Elfimov_et_al:2008}. 
(The calculated YMnO$_3$ DOS shown in Fig.~\ref{ymo_structure} (c) was also obtained using a $U_{eff}$ of 7 eV on the Mn-$d$ states, 
consistent with literature studies on YMnO$_3$.)
A $10\times 1 0\times 4$ Gamma-centred k-point mesh was used for Brillouin-zone integrations. The plane-wave cut-off was set to 550eV and in performing the structural optimizations, we allowed the ions to relax until the Hellmann-Feynman forces were less than 1 meV/\AA$^{-1}$. 

To treat the electron-electron correlations in modeling the metal-insulator transition, we used the Dynamical Mean-Field Theory (DMFT) 
approach\cite{Georges_et_al:1996}. We first constructed maximally localized Wanner functions for the transition-metal $d$ bands of the 
Kohn-Sham Hamiltonian using the wannier90 code\cite{Marzari:2012, Mostofi:2008}. We then used these as the noninteracting part $H_{0}$ of 
a Hubbard Hamiltonian $H=H_{0}+H_{int}$ where $H_{int}$ is the local electron-electron interaction comprising the intra-orbital Coulomb 
interaction, $U_{DMFT}$, and Hund's rule coupling, $J$. Here we used $J=0$ eV since we consider a single orbital, while we varied 
$U_{DMFT}$ from 0 to 4 eV. We used a continuous-time hybridization expansion quantum Monte Carlo solver\cite{Gull:2011} as implemented 
in the TRIQS 1.0 code\cite{triqs} to calculate the local Green's function within DMFT at a temperature $T=1/(k_{B}\beta) = 290$K 
($\beta = 40$ eV$^{-1}$).

\section{Results}

\subsection{Structural optimization}
First we perform an initial band structure screening of our trial compounds within the high-temperature $P6_{3}/mmc$ structure. Our 
calculated lattice constants for all six compounds are given in Table~\ref{structure_table}. 
For the oxides and fluorides, the transition metal - apical ligand distances are shorter than the corresponding in-plane distances,
which should yield the highest energy singly degenerate crystal field state required in our analysis. 
As expected from the larger atomic radius of Na, the volume of NaCuF$_3$ is slightly greater than that of LiCuF$_3$, although
the apical Cu-F bonding distances are almost the same in both compounds. Between the Zr and Sn oxides, the volumes and Cu-O 
distances are almost identical, suggesting similar crystal-field splittings in the two cases. Interestingly, for the sulphides,
the apical Cu-S bonding distances are longer than the in-plane distances. This should place the singly degenerate $d_z^2$
band in the lowest energy position which will lead to its full application. Therefore we exclude the sulphides from our
list of candidates for further consideration. 

\subsection{Electronic properties at the density functional theory level}
\begin{figure*}
 \centering
 \includegraphics[width=10cm, keepaspectratio=true]{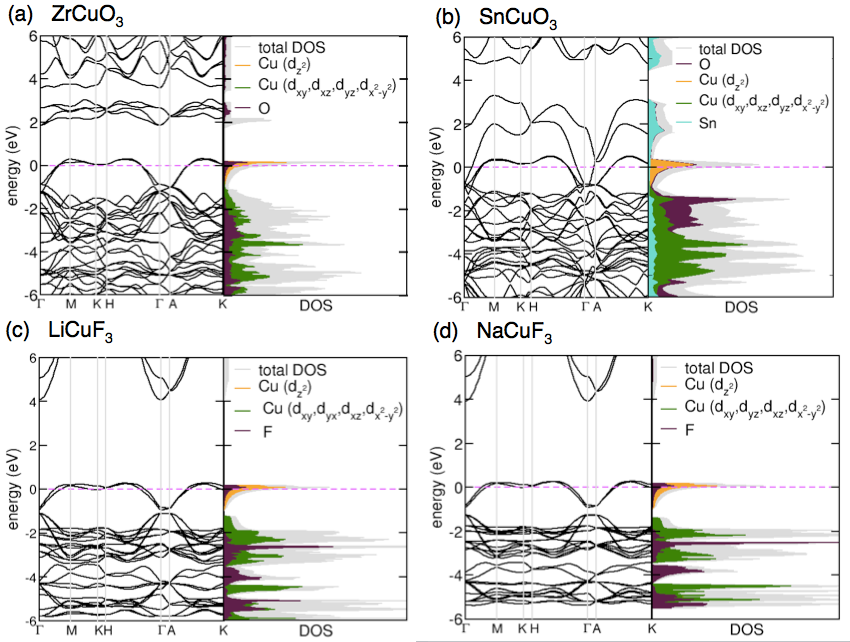}
 \caption{Calculated band structures and total and orbital-projected densities of states (DOS) for 
(a) ZrCuO$_3$, (b) SnCuO$_{3}$, (c) LiCuF$_{3}$ and NaCuF$_{3}$, all in the hexagonal manganite structure. The Fermi level 
is set to $0$ eV and is marked by the dashed line in each of the plots.}
 \label{alldos}
\end{figure*}

Using our optimized structures, we next calculate the electronic properties of the candidate materials at the density functional theory level.
Fig.~\ref{alldos} shows the calculated electronic band structures and orbital-resolved densities of states for (a) ZrCuO$_3$, (b) SnCuO$_3$, 
(c) LiCuF$_{3}$ and (d) NaCuF$_{3}$. 
For both of the oxide compounds, the Fermi energy intersects a band that is composed primarily of Cu-$d_{z^2}$ states, with some admixture
from apical O-$p_z$ states. The Cu-$d_{z^2}$ band is almost separated from the broad valence band composed of the other four majority Cu-$d$ 
bands and oxygen, but the separation is not complete due to its non-negligible bandwidth of 1.2 eV. 
For the tin case, we also find Sn-$s$ states at E$_{F}$ (Fig.~\ref{alldos} (b)), which are undesirable for achieving a single-band model. 

We find a more promising situation in the fluorides (Figs.~\ref{alldos} (c) and (d)) both of which show a completely split-off 
half-filled Cu-$d_{z^2}$ band. The reduced anion hybridization in these more ionic compounds gives a smaller bandwidth of $\sim$ 1eV and makes
the fluorides more single-band-like. 
(Note that our calculation is for a two-formula-unit unit cell, and the two Cu ions are slightly symmetry inequivalent so two Cu $d_z^2$ bands are
shown.)  
The electronic structures of LiCuF$_3$ and NaCuF$_3$ are remarkably similar in the region surrounding the Fermi level since there is no contribution 
from the A sites in either case in this region. The only clear difference is that the crystal-field gap between the split-off band and the other 
Cu-$d$ states is slightly larger in NaCuF$_3$ ($\sim$ 0.3eV) than in LiCuF$_3$ ($\sim$0.2 eV).

In summary, we find from density functional calculations that the cuprates of Zn and Sn, and the copper fluorides of Li and Na are promising
candidate single-band Hubbard model materials when in the hexagonal manganite structure, with the fluorides having the more desirable
properties. Next we continue with one of the fluoride representatives  - LiCuF$_{3}$ - and investigate the effect of explicit electron correlations 
at the dynamical mean field theory level. 

\subsection{Electronic structure at the dynamical mean field theory level}
\begin{figure*}
 \centering
 \includegraphics[width=8cm, keepaspectratio=true]{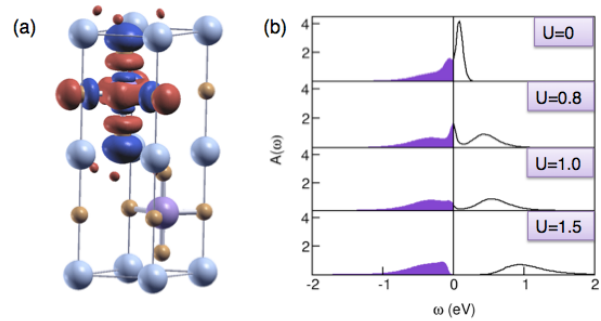}
 \caption{(a) Maximally localized Wannier function in LiCuF$_3$ obtained from projection of the $d_{z^{2}}$ band for one Cu site. 
(b) Calculated DMFT spectral function for a range of $U$ values as a function of frequency $\omega$. The Fermi energy is indicated
to 0 eV, and filled orbitals are shaded.}
 \label{dmft}
\end{figure*}

To study the effect of explicit electron-electron correlations on the d$_z^{2}$ band, we next calculate the spectrum of LiCuF$_{3}$ 
using dynamical mean field theory (DMFT). We first construct the maximally-localized Wannier function (Fig.~\ref{dmft} (a)) on a Cu site 
for the energy window corresponding to the isolated Cu-$d_{z^{2}}$ band using the Wannier90 code\cite{Mostofi:2008}. 
As expected the Wannier function has primarily $d_{z^{2}}$ character with small $p$-symmetry ``tails'' extending onto the neighboring fluorine 
ions. 

Using this Wannier function as a basis, we next calculate the spectral function, $A(\omega)$ within DMFT,
by analytic continuation using the maximum 
entropy method\cite{mem} Our results are shown in Fig.~\ref{dmft}(b) for $U_{DMFT}$ values ranging from 0 to 1.5 eV. 
We see that the system is metallic at $U=0$ eV. As the interaction parameter $U$ is increased, the spectral weight gradually shifts from the 
quasiparticle peak at $\omega$ = 0 eV to the upper and lower Hubbard bands. Eventually a classic Mott transition occurs when the electron interactions 
are strong enough to cause the quasiparticle peak to disappear completely and a gap to form. Consistent with our Tr $G(\beta/2)$ data (not shown), 
the system gaps above $U_{DMFT}=1.25$ eV.
We comment briefly on the apparent inconsistency introduced by our use of a non-zero $U$ value in our original DFT calculations, combined
with a second $U$ in the DMFT study. To address this issue, we repeated our DFT calculations in the GGA ($U=0$) limit and found the band
structure and band widths to be very similar to the GGA$+U$ case, with the DFT$+U$ treatment acting as a 'scissors operator' to isolate the 
$d_{z^{2}}$ orbital. Therefore we expect only small quantitative changes in the $U_{DMFT}$ value at the metal-insulator transition if the 
initial Wannier functions were constructed from results of the $U=0$ GGA calculation; construction of the Wannier functions would be 
complicated, however, by the poorly isolated bands.

\subsection{Structural stability}
As a first step to assessing the feasibility of synthesizing our proposed compounds, we calculate the relative structural stabilities
of a range of structural isomorphs of LiCuF$_3$. We consider the following crystal structures all of which 
are known to exist for compounds with $ABX_{3}$ chemistry:
high-temperature hexagonal manganite ($P6_{3}/mmc$), low-temperature ferroelectric hexagonal manganite ($P6_{3}cm$), low-temperature 
nonpolar hexagonal manganite ($P\bar{3}c$), cubic perovskite ($Pm\bar{3}m$), ilmenite ($R\bar{3}$), the `NaCuF$_3$' structure (which 
corresponds to the ambient experimental ground state, $P\bar{1}$), the `CoGeO$_3$' structure ($C2/c$), and two non-centrosymmetric
structural variants -- rhombohedral ($R3m$) and tetragonal ($P4/mmm$) -- found in BaTiO$_3$.
We apply both positive and negative hydrostatic pressure by scaling the lattice parameters, keeping the fractional internal coordinates 
of the ions fixed for each calculation to ensure that the original symmetry group is maintained. 

\begin{figure*}
 \centering
 \includegraphics[width=8cm, keepaspectratio=true]{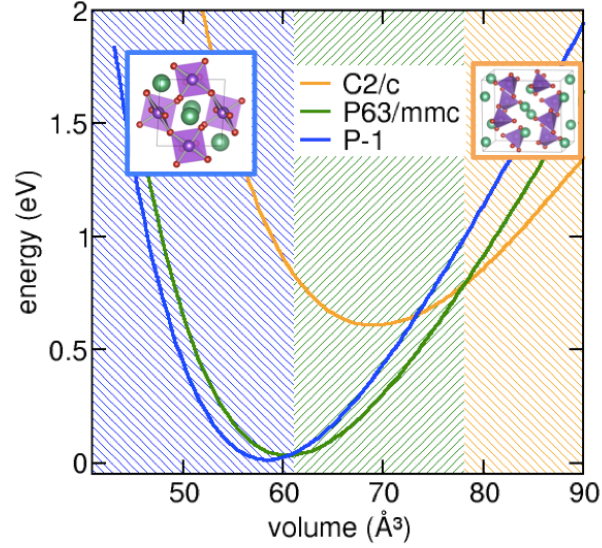}
 \caption{Calculated total energy as a function of volume for the three calculated lowest-energy LiCuF$_3$ isomorphs. 
The NaCuF$_3$ structure (blue shaded region and inset) is the most stable structure for small volumes, and the hexagonal manganite becomes
the lowest energy structure for volumes greater than 63\AA$^3$ (green shaded region). Finally for very large volumes the CoGeO$_3$ structure is stabilized (shown in orange inset).}
 \label{volume_li}
\end{figure*}

Fig.~\ref{volume_li} shows the resulting calculated energies as a function of volume for the various structural isomorphs of LiCuF$_3$ that were
lowest in energy over the range of volumes studied. 
We find that the lowest energy structure is the `NaCuF$_3$' structure, however for an 8 \% increase in volume (corresponding to a 2 \%
increase in lattice constant) the hexagonal manganite structure is the ground state, suggesting that it might be possible to achieve
it using tensile strain. At much larger lattice constants the $C2/c$ CoGeO$_3$ structure becomes stable.

\subsection{Doping}
Finally, we explore routes to electron- and hole- doping LiCuF$_3$ so that the entire range of band filling can be accessed. 
First we artificially add (remove) up to one electron per formula unit without changing the ion configuration, using a compensating 
background positive (negative) charge to prevent electrostatic divergence. As expected, the Fermi level shifts downwards (upwards)
as electrons are removed (added), giving a completely unfilled Cu-$d_{z^2}$ band when 1 e$^-$ per formula unit is removed, and a
completely filled Cu-$d_{z^2}$ band when 1 e$^-$ per formula unit is added. No qualitative change apart from the rigid shift in the
Fermi level is found over the entire doping range; in particular the split-off single-band character is retained. 

\begin{figure*}
 \centering
 \includegraphics[width=15.8cm, keepaspectratio=true]{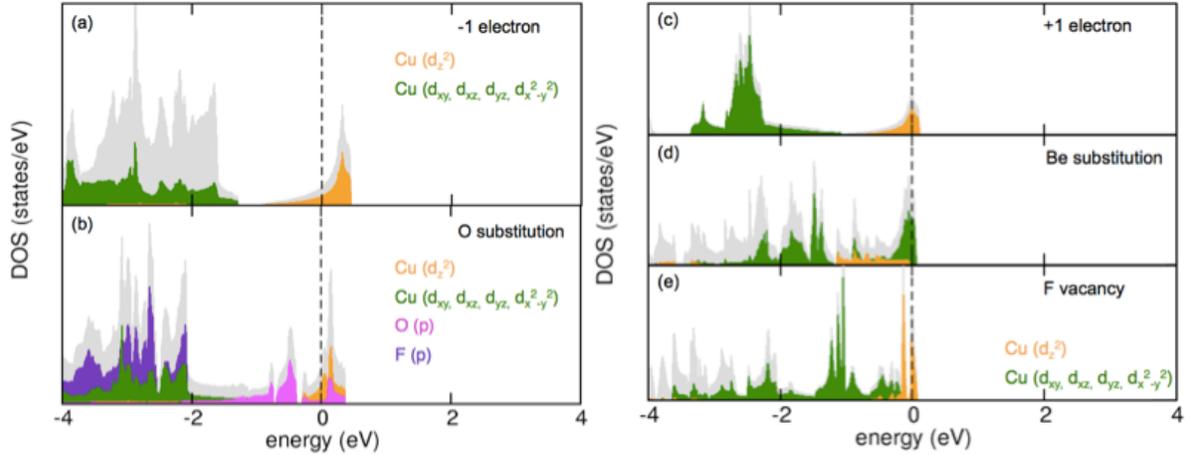}
 \caption{Left. a) Density of states of LiCuF$_{3}$ with half an electron removed per formula unit. b) Density of states for 
oxygen-substituted Li$_2$Cu$_2$F$_5$O$_1$. 
Right. (c) Density of states of LiCuF$_{3}$ with half an electron added per formula unit.  (d) 
Calculated DOS for Be-substituted LiCuF$_3$. (e) Calculated DOS for  LiCuF$_3$ with a fluorine vacancy. In all cases the Fermi level 
is set to $0$eV in each case, and orbital-projected DOSs are shown for the relevant atoms.}
\label{licuf3_fdoping}
\end{figure*}

An experimentally accessible possibility for hole doping is to substitute fluorine (which forms F$^-$, with one negative charge per atom)
by oxygen (which forms O$^{2-}$, with two). Replacing one fluorine atom per unit cell with oxygen yields the chemical formula 
Li$_{2}$Cu$_{2}$F$_{5}$O$_{1}$, which corresponds to removing half an electron per formula unit. We calculated the relative energies of 
the substitutional sites for the oxygen by replacing each of the two inequivalent fluorine ions 
with an oxygen ion, performing a full structural relaxation, then comparing the final energies. We found that in-plane subsitution
is lower in energy by $\sim$0.3 eV per formula unit than apical substitution; this is favorable for our Hubbard material design as the in-plane ions have
minimal hybridization with the split-off single band. The resulting density of states (DOS) for in-plane oxygen substitution 
is plotted in Fig.~\ref{licuf3_fdoping}(b) (left panel), with the orbital projected O-$p$ states shown in pink. Note that the electronic
structure in the region of the Fermi level is remarkably similar to that obtained by removing electrons in our first doping 
calculations. 

Possible routes to electron doping LiCuF$_3$ are to replace some Li with Be or to introduce F vacancies. Substituting one Li ion with a Be 
ion introduces half an extra electron per formula unit with the chemical formula LiBeCu$_{2}$F$_{6}$. The influence of Be substitution on 
the electronic structure is shown in Fig.~\ref{licuf3_fdoping}(b) (right panel), where we have plotted the Cu-$d_{z^2}$ band separately 
from the other Cu-$d$ states. While no Be states appear near the Fermi level, the substitution causes an orbital reordering whereby the 
region next to the Fermi level now comprises contributions from the full $d$ manifold. 
Removing one fluorine atom per unit cell to give the chemical formula Li$_{2}$Cu$_{2}$F$_{5}$ also corresponds to adding half an electron 
per formula unit. We calculated the relative energy of vacancy sites by removing the fluorine ion from each of the two inequivalent fluorine 
in turn and performing a full internal relaxation. We found that apical fluorine vacancies are lower in energy than in-plane vacancies by 
$\sim$0.1 eV per formula unit; the resulting density of states for apical vacancies is plotted in Fig.~\ref{licuf3_fdoping}(e) (right panel). 
Unlike the Be substitution case, the region surrounding the Fermi level retains its Cu-$d_{z^2}$ character and behaves more like the rigid-band 
model of electron addition.

\subsection{Exotic superconductivity}

Since their discovery in the 1980's\cite{Bednorz/Muller:1986}, the cuprate superconductors have remained one of the most studied classes 
of materials, as well as one of the most elusive in revealing a comprehensive understanding of their behavior\cite{Anderson:1997, Leggett:2006}. 
In particular, while 
they have the highest critical temperatures of all known superconductors, the details of the pairing mechanism are still unknown. 
The relevant features for high-$T_{c}$ superconductivity in the cuprates appear to be the quasi two-dimensional electronic structure, 
the parent antiferromagnetic compound with a range of accessible dopings, the spin-$\frac{1}{2}$ magnetic moment and the single band 
at the Fermi level.
For example, the La$_{2}$CuO$_{4}$ cuprate parent compound contains 3$d^{9}$ octahedrally coordinated Cu$^{2+}$ ions. A strong Jahn-Teller
distortion lifts the energy of the two $e_g$ orbitals yielding a non-degenerate half-filled Cu-$d_{x^{2}-y^{2}}$ band which is, however, 
strongly hybridized with the $p_{x}$ and $p_{y}$ states on the neighboring oxygens\cite{Zhang/Rice:1988}. The ground state is antiferromagnetic 
and insulating due to the combined effects of exchange splitting and Mott electron repulsion. 
Recognition of these characteristics have led to attempts to design new materials that might enhance exotic superconductivity. A particularly
clever suggestion was to use geometric engineering to crystal-field split the degenerate $e_g$ orbitals rather than relying on the
Jahn-Teller distortion\cite{Chaloupka/Khaliullin:2008}. A proposed candidate system was a layered superlattice of (LaNiO$_3$/LaAlO$_3$) 
in which the symmetry and strain of the superlattice would lift the degeneracy of the Ni \textit{e$_g$} states. While an intriguing
idea, in experiment charge-transfer physics has been found to dominate the behavior and supress superconductivity. 

Our bespoke single-band Hubbard material, LiCuF$_3$, exhibits all the necessary ingredients for exotic superconductivity. In particular, 
the single band at the Fermi level is strongly two-dimensional, and Cu$^{2+}$ ions are spin-$\frac{1}{2}$ with antiferromagnetic
interactions. The triangular lattice might lead to additional exotic behavior. 
Therefore, as a final analysis, we investigate whether a superconducting transition might occur in this compound and provide a prediction 
of its critical 
temperature. For this we use a variant of the Dynamical Cluster Approximation (DCA) \cite{Jarrell_et_al:2001}, which is the cluster extension of the 
(DMFT)\cite{Georges_et_al:1996}. In this method, the original infinite lattice problem is transformed into solution of a finite-sized cluster 
with periodic boundary conditions, embedded in a self-consistent mean field. This transformation is achieved via a coarse-graining procedure 
of the Green's function, in which the Brillouin zone is divided into $Nc$ patches on which the self-energy is assumed to be constant. 
The DCA is able to treat all short range correlations between the electrons in the cluster exactly, while long-range 
correlations outside the cluster are taken into account by the embedding self-consistent mean-field. 
We recently extended the method\cite{Staar/Maier/Schulthess:2013,Staar_et_al:2013,Staar/Maier/Schulthess:2014}, to allow determination of 
the critical temperature for 
large clusters at a strong interaction strength and to treat continuous lattice self energies and vertex functions in momentum space, 
giving a more accurate description and allowing a more stringent investigation of the superconducting gap function in 
momentum space\cite{Staar/Maier/Schulthess:2014}. We use this new DCA+ method here.

 DCA has previously been used to investigate the pairing mechanism of the Cwoper pairs in the high-$T_c$ cuprates \cite{PhysRevLett.95.237001,PhysRevB.84.180513,PhysRevB.84.220506,PhysRevLett.110.216405,PhysRevB.89.195133,PhysRevB.90.041110,PhysRevB.91.085116}. 
In addition, Chen et al.\cite{Chen_et_al:2013} recently investigated the superconducting transition on a triangular lattice and discovered 
two divergent pairing susceptibilities, one with $d_{xy}$ symmetry and one with $d_{x^2-y^2}$ symmetry, for 
a cluster of size 6 and $U = 8.5\,t$. 
Furthermore, their phase diagram suggests that the critical temperature 
rises as the doping is decreased, up to a possible $T_c\approx0.06\,t$ (with t being the hopping parameter). 
This further motivates our study, as our LiCuF$_3$ compound has a half-filled band.

\begin{figure*}
 \centering
 \includegraphics[width=8cm, keepaspectratio=true]{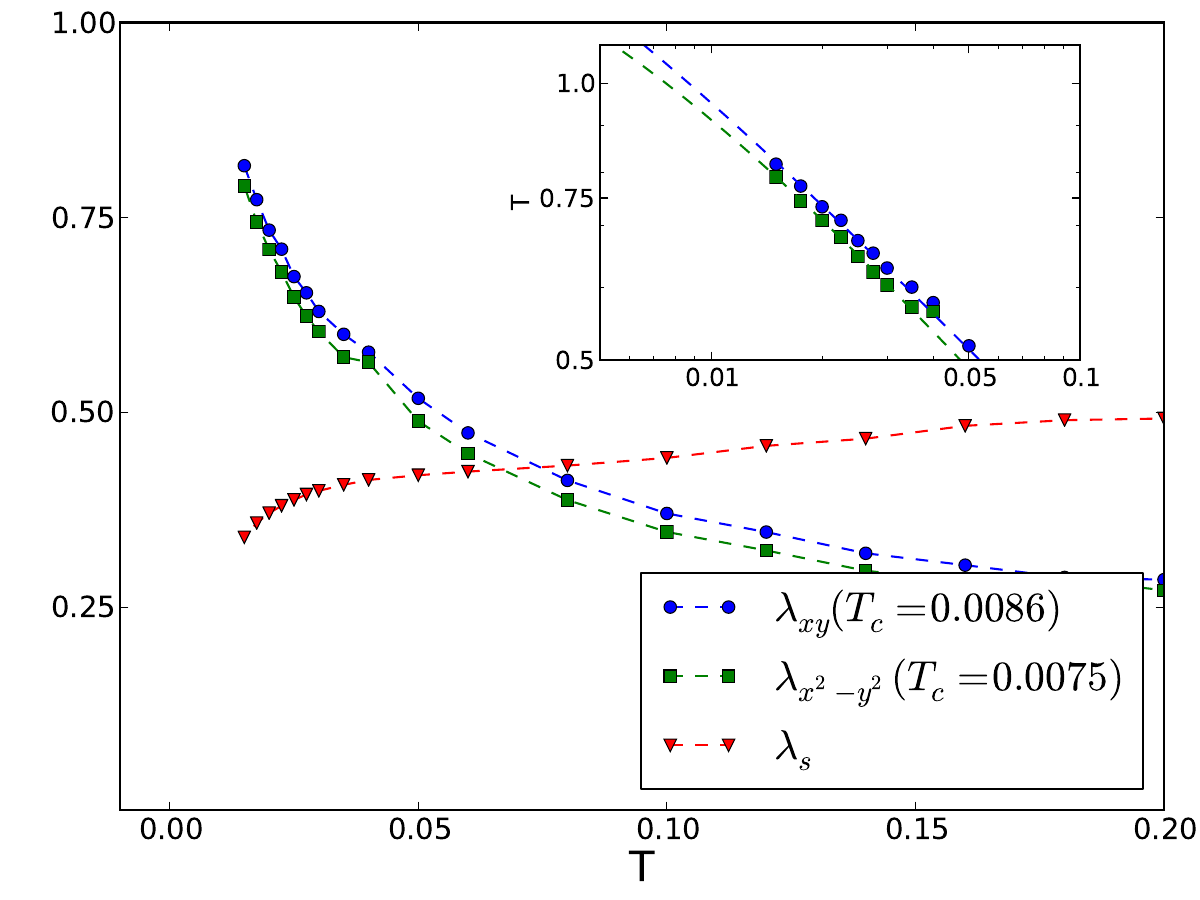}
 \caption{The leading eigenvalue $\lambda$ of the matrix $\Gamma\,\chi$. A superconducting transition occurs when this leading eigenvalue $\lambda$ crosses 1 since the pairing susceptibility $\chi = \chi_0/(1-\Gamma\,\chi)$ will diverge. Inset: A log-log plot of the leading eigenvalue versus temperature. The linearity of the plot suggests that $\lambda$ behaves as $\lambda \propto \alpha\,(T-T_c)^\gamma$. This behaviour indicates that we are in the mean field regime. This allows us to extrapolate lambda and find a $T_c$ equal to $\propto 0.08$.}
 \label{Staar_1}
\end{figure*}

\begin{figure*}
 \centering
 \includegraphics[width=16cm, keepaspectratio=true]{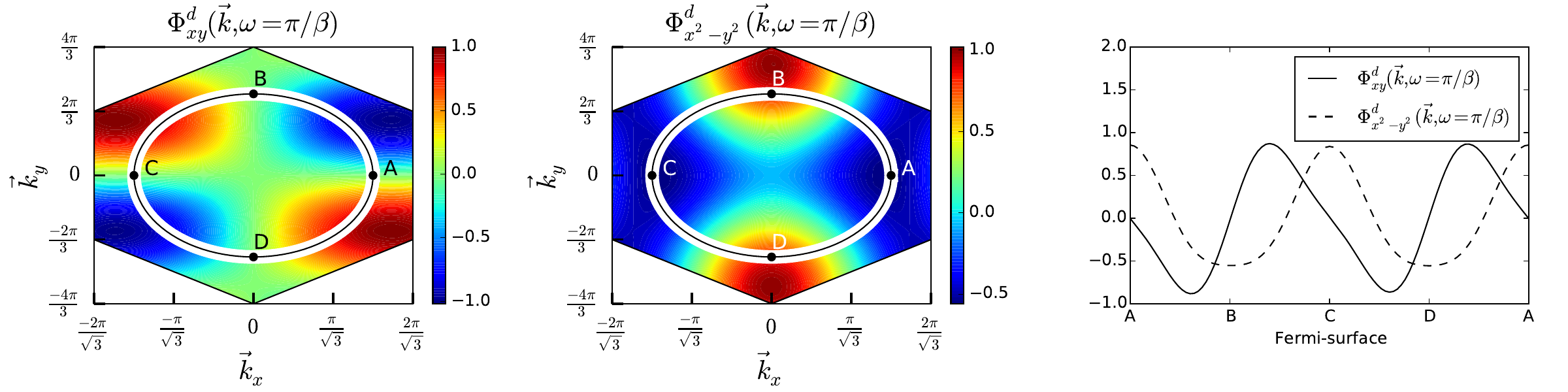}
 \caption{The momentum space structure of the leading superconducting gap functions $\Phi^{d_{x\,y}}$ (left) and $\Phi^{d_{x^2-y2}}$ (middle) at the lowest Matsubara frequency. By looking at their projections on the Fermi-surface (right), we can understand why the $\Phi^{d_{x^2-y2}}$ has a lower critical temperature. Its gap is simply not as big as the gap of  $\Phi^{d_{x\,y}}$ in the negative region. Since the critical temperature is proportional to the size of the gap, one would expect a lower $T_c$ for $\Phi^{d_{x^2-y2}}$ than for $\Phi^{d_{x\,y}}$.}
 \label{Staar_2}
\end{figure*}

We use the DCA+ method at $U=8t$ to solve a 16-site cluster, which has been used in earlier 
papers investigating the high-$T_c$ cuprate superconductors to allow for a direct comparison with these earlier calculations. 
In Fig.~\ref{Staar_1}, we show our calculated leading eigenvalue $\lambda$ of the matrix $\Gamma\,\chi$. 
If this leading eigenvalue $\lambda$ crosses 1, the pairing susceptibility $\chi = \chi_0/(1-\Gamma\,\chi)$ diverges and a superconducting 
transition occurs. As in Ref.~\onlinecite{Chen_et_al:2013}, we find two diverging superconducting modes, one with $d_{xy}$ and one 
with $d_{x^2-y^2}$ symmetry. In the inset, we show our finding that the leading eigenvalues are linear on a log-log plot, which indicates 
that the eigenvalue behaves as $\lambda \propto \alpha\,(T-T_c)^\gamma $ and is thus in the mean-field regime. This allows us to extrapolate 
to lower temperatures and find a $T_c\approx0.08$ by solving the equation $\lambda=1$. In order to understand why the $d_{xy}$ superconducting 
mode is favoured over the $d_{x^2-y^2}$, one can look at their corresponding gap functions $\Phi$, defined as the corresponding 
eigenvector, in momentum space. 
In Fig.~\ref{Staar_2} , we show the momentum-space structure of the gap function at the lowest 
matsubara frequency for respectively the $d_{xy}$ and the $d_{x^2-y^2}$ superconducting modes as well as a cut of the superconducting gaps 
along the Fermi surface. We see that the amplitude of the $d_{xy}$ mode is larger than that of the $d_{x^2-y^2}$ mode, which leads to a 
larger lambda and critical temperature.

\section{Summary}
We have identified a class of materials with the hexagonal manganite structure whose low-energy behavior reflects the single-band
Hubbard Hamiltonian, and propose LiCuF$_{3}$ and NaCuF$_{3}$ as promising candidates. We used density functional
theory, dynamical mean field theory and the dynamical cluster approximation to characterize LiCuF$_3$ in detail and showed that it displays
the expected Mott transition, as well as exotic $d$-wave superconductivity. 
We hope that our work will motivate experimental synthesis and characterization of this and related compounds
\cite{Ylvisaker/Pickett:2006,Lee_et_al:2007}, as well as exploration of bespoke materials with other model Hamiltonians.

\begin{acknowledgments}

This work was supported by the ETH Z\"urich, the ERC Advanced Grant program, No. 291151 and the NCCR MARVEL.
We thank Gabriele Sclauzero, Claude Ederer, Krzysztof Dymkowski and Sverre M. Selbach for helpful discussions. 
Calculations were performed at the CSCS Computing Facility.
\end{acknowledgments}

The authors declare that they have no competing financial interests.

Correspondence and requests for materials should be addressed to nicola.spaldin@mat.ethz.ch 

\clearpage

\begin{table*}

\begin{tabular}{|l||c|c|c|c|}
\hline
       & \begin{tabular}[c]{@{}c@{}}\textbf{a}\\ (\AA) \end{tabular} & \begin{tabular}[c]{@{}c@{}}\textbf{c}\\ (\AA)\end{tabular} & \begin{tabular}[c]{@{}c@{}}\textbf{Cu-F} (in plane)\\ (\AA)\end{tabular} & \begin{tabular}[c]{@{}c@{}}\textbf{Cu-F} (apical)\\ (\AA)\end{tabular} \\ \hline
ZrCuO$_{3}$ & 3.75                                          & 9.33                                         & 2.16                                                        & 1.81                                                      \\ \hline
SnCuO$_{3}$ & 3.78                                          & 9.38                                         & 2.18                                                        & 1.82                                                      \\ \hline
LiCuF$_{3}$ & 3.47                                          & 11.13                                         & 2.00                                                        & 1.85                                                      \\ \hline
NaCuF$_{3}$ & 3.60                                          & 11.60                                         & 2.08                                                        & 1.84                                                      \\ \hline
ZrCuS$_{3}$ & 4.69                                          & 10.26                                         & 2.70                                                        & 2.22                                                      \\ \hline
SnCuS$_{3}$ & 4.72                                          & 10.94                                        & 2.71                                                        & 2.36                                                     \\ \hline
\end{tabular}
\caption{Calculated structural parameters for hypothetical oxides, fluorides and sulphides in the centrosymmetric hexagonal manganite 
structure with the $P6_{3}/mmc$ space group. The lattice parameters, $a$ and $c$ are given in the standard hexagonal setting with two 
formula units per unit cell.}
\label{structure_table}
\end{table*}

\end{document}